\documentclass[osajnl,twocolumn,showpacs,superscriptaddress,10pt]{revtex4-1} 
\usepackage{amsmath,amssymb,graphicx}
\begin{document}

\title{Non-invasive monitoring and control in silicon photonics by CMOS integrated electronics}

\author{Stefano Grillanda}\email{Corresponding author: stefano.grillanda@polimi.it}
\affiliation{Dipartimento di Elettronica, Informazione e Bioingegneria, Politecnico di Milano, 20133 Milano, Italy}

\author{Marco Carminati}\thanks{S.G. and M.C. contributed equally to this work.}
\affiliation{Dipartimento di Elettronica, Informazione e Bioingegneria, Politecnico di Milano, 20133 Milano, Italy}

\author{Francesco Morichetti}
\affiliation{Dipartimento di Elettronica, Informazione e Bioingegneria, Politecnico di Milano, 20133 Milano, Italy}

\author{Pietro Ciccarella}
\affiliation{Dipartimento di Elettronica, Informazione e Bioingegneria, Politecnico di Milano, 20133 Milano, Italy}

\author{Andrea Annoni}
\affiliation{Dipartimento di Elettronica, Informazione e Bioingegneria, Politecnico di Milano, 20133 Milano, Italy}

\author{Giorgio Ferrari}
\affiliation{Dipartimento di Elettronica, Informazione e Bioingegneria, Politecnico di Milano, 20133 Milano, Italy}

\author{Michael Strain}
\affiliation{Institute of Photonics, The University of Strathclyde, Glasgow G4 0NW, UK}

\author{Marc Sorel}
\affiliation{School of Engineering, University of Glasgow, Glasgow, G12 8QQ, UK}

\author{Marco Sampietro}
\affiliation{Dipartimento di Elettronica, Informazione e Bioingegneria, Politecnico di Milano, 20133 Milano, Italy}

\author{Andrea Melloni}
\affiliation{Dipartimento di Elettronica, Informazione e Bioingegneria, Politecnico di Milano, 20133 Milano, Italy}

\begin{abstract} As photonics breaks away from today's device level toward large scale of integration and complex systems-on-a-chip, concepts like monitoring, control and stabilization of photonic integrated circuits emerge as new paradigms. Here, we show non-invasive monitoring and feedback control of high quality factor silicon photonics resonators assisted by a transparent light detector directly integrated inside the cavity. Control operations are entirely managed by a CMOS microelectronic circuit, hosting many parallel electronic read-out channels, that is bridged to the silicon photonics chip. Advanced functionalities, such as wavelength tuning, locking, labeling and swapping are demonstrated. The non-invasive nature of the transparent monitor and the scalability of the CMOS read-out system offer a viable solution for the control of arbitrarily reconfigurable photonic integrated circuits aggregating many components on a single chip. \end{abstract}



\maketitle 

\section{Introduction}

The level of complexity achieved by today's integrated electronic systems is commonly perceived as the result, primarily, of challenging technological efforts to scale device dimensions down to their ultimate physical limits \cite{Chau_2007_NatureMat}. Although this is indeed true, it is only partially responsible for their success, as miniaturization is not synonymous with large scale integration. In fact analog electronic circuits cannot function properly without adequate tools to dynamically steer and hold each embedded device to the desired working point, counteracting functional drifts due to fluctuations in the environment, aging effects, mutual crosstalk, and fault events \cite{Gonzalez_1997_JSSC}. 

This argument is directly applicable to photonic technologies. Even though photonic platforms, like silicon-on-insulator (SOI), have demonstrated the maturity for squeezing several thousands of photonic elements in a footprint of less than 1 mm$^2$ \cite{Sun_2013_Nature}, the  evolution of integrated photonics from device level to large scale systems is still a challenge. In fact, when aggregating several components on a single chip, the aforementioned parasitic effects become critical and need to be addressed through feedback control loops that locally monitor and continuously set each optical element to the desired functionality \cite{Morichetti_2014_JPHOT}.

These issues are particularly severe in Si photonics microresonators, because of their extreme sensitivity to fabrication tolerances \cite{Xia_2007_NatPhoton} and temperature variations \cite{Guha_2010_OpEx}. Several approaches have been recently proposed to lock the resonant wavelength of Si microrings, for instance  by applying dithering signals \cite{Padmaraju_2014_JLT} or homodyne detection schemes \cite{Cox_2014_OpEx}, and by monitoring the power level \cite{Padmaraju_2012_OpEx,Padmaraju_2013_OpEx} or the bit-error-rate \cite{Zortman_2013_IEEEMicro} of the optical signal. However, all the techniques proposed so far require the use of on-chip or external photodetectors to partially tap the light travelling in the resonator. Though effective on single devices, this approach is not scalable to large scale integration circuits \cite{Baehr-Jones_2012_NatPhoton}, where multi-point light tapping would incur in a large amount of optical power wasted for monitoring operation. Local feedback control assisted by transparent optical detectors is envisioned as an enabling tool for the realization of complex and arbitrarily reconfigurable systems-on-a-chip \cite{Miller_2014_OPN,Miller_2013_PhotonRes,Doylend_2012_LPR}. 


Here, we show a Si photonic-electronic integrated platform enabling the feedback control of Si photonics integrated circuits without the need of tapping any photon from the waveguide. The status of high quality factor resonators is monitored by a recently pionereed ContactLess Integrated Photonic Probe (CLIPP) \cite{Morichetti_2014_CLIPP}, that realizes a fully transparent detector and can be integrated directly inside any photonic circuit, including microrings. The feedback loop, combining the CLIPP read-out system and the microring control functions, is entirely integrated onto an electronic CMOS circuit \cite{Carminati_2014_Impedimetric} that is wire-bonded to the Si photonics chip. Advanced functionalities and control operations, such as wavelength tuning, locking, labeling and swapping are demonstrated in a thermally actuated resonator, proving that the presented Si photonic-electronic integrated platform is an efficient and flexible solution for the realization and control of Si photonics circuits hosting many components.

\begin{figure}[t]
\centering\includegraphics[width=8.3cm]{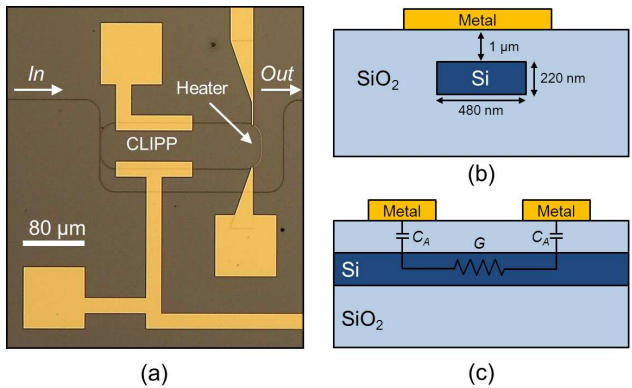}
\caption{(a) Top-view photograph of the fabricated Si microring, where CLIPP and thermal actuator are integrated. (b) Cross section of the Si core waveguide, with the CLIPP metal electrode deposited on top of the SiO$_2$ cladding. (c) Longitudinal profile of the Si waveguide showing the CLIPP equivalent circuit in the electrical domain, that is composed of two access capacitances \textit{C$_A$} and a resistor of conductance \textit{G}.}
\label{fig:Fig0}
\end{figure}

\begin{figure*}[t]
\centering\includegraphics[width=17.7cm]{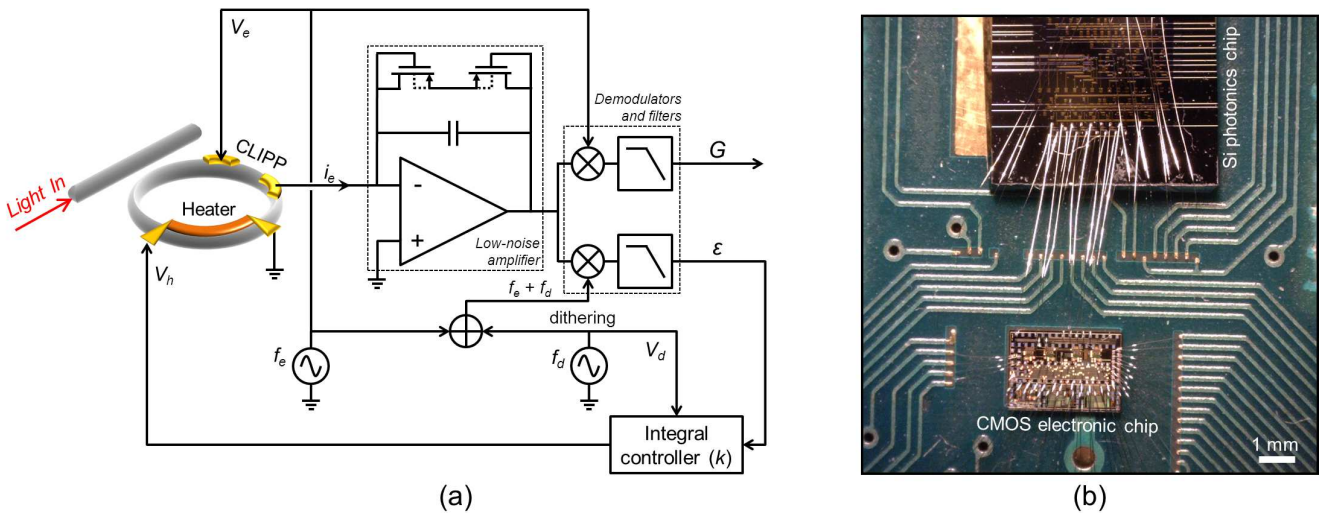}
\caption{(a) Schematic of the electronic integrated circuit that performs the read-out of the CLIPP electric signal, and manages control operations of the Si photonics microring resonator (such as wavelength tuning, locking, labeling and swapping). (b) Photograph of the Si photonics chip (on the top) hosting the microring resonator, that is wire-bonded to the CMOS electronic chip (in the bottom) containing all the CLIPP read-out and microring control circuitry. Both the photonic and the electronic chips are integrated onto the same printed circuit board.}
\label{fig:Fig1}
\end{figure*}

\section{CLIPP concept and fabrication}

Figure \ref{fig:Fig0}(a) shows a top-view photograph of a microring fabricated in silicon-on-insulator (SOI) technology, where the CLIPP electrodes and the thermal actuator are integrated.
 The microring is 516 $\mu$m long, has 20 $\mu$m bending radius, and is realized by a channel waveguide with 480 nm wide and 220 nm thick Si core [Fig. \ref{fig:Fig0}(b)], patterned by means of electron-beam lithography \cite{Gnan_2008_ElnLett}. The waveguide core is buried into a 1 $\mu$m thick silicon dioxide (SiO$_2$) top cladding, that is grown by plasma enhanced chemical vapor deposition (PECVD) \cite{Morichetti_2014_CLIPP,Gnan_2008_ElnLett}. On top of it, the metallic NiCr heaters and Au pads of the CLIPP are patterned by lift-off technique, and placed at sufficient distance from the Si core to avoid any significant absorption of the optical mode by the metal.   

The CLIPP is constituted simply by the two 100 $\mu$m long metal electrodes placed on top of the microring waveguide, here mutually spaced by about 83 $\mu$m on the side of the bus-to-ring directional coupler. The CLIPP can be fabricated by using any CMOS-compatible metal technology, and can exploit traditional processes used for the fabrication of thermal actuators, without requiring any additional or specific process step.

The equivalent electrical circuit of the CLIPP is reported in Fig. \ref{fig:Fig0}(c) that shows the longitudinal profile of the waveguide (stray capacitive coupling to the Si substrate is here neglected). Due to the typical doping of SOI wafers (10$^{15}$ cm$^{-3}$, p-type) the Si core acts mainly as a resistor of conductance $G$, whereas the insulating top cladding provides the access capacitances $C_{A}$. The CLIPP monitors variations of the waveguide electric conductance $\Delta G$ with optical power $ P $, that are induced by a carrier generation effect, occurring at the native Si/SiO$_2$ interface, associated to intrinsic surface-state-absoption (SSA) processes \cite{Morichetti_2014_CLIPP, Baehr-Jones_2008_OpEx}. These phenomena exist even in ideally smooth interfaces because of the termination of the Si lattice at the walls of the waveguide core \cite{Monch_2001_Book}. No specific treatment is performed on the Si core surface.

The CLIPP is essentially a non-invasive observer of the local power propagating in an optical waveguide \cite{Morichetti_2014_CLIPP}, and therefore can be easily utilized in any applications requiring to monitor the light travelling in a photonic circuit (see Sec. \ref{sec:tuning}, \ref{sec:locking}, and \ref{sec:swapping}). In addition, it is worth noticing that, thanks to its intrinsic non-perturbative nature, the CLIPP can be placed in any point of the circuit to monitor its local status without affecting its functionality. 


\section{CLIPP read-out system}

The CLIPP observes directly the amount of light stored in the resonant cavity, information that traditionally is not available unless a portion of the optical power is tapped and rerouted to a photodetector, by measuring variations of the electric conductance $G$ of the waveguide core \cite{Morichetti_2014_CLIPP}.

\begin{figure}[t]
\centering\includegraphics[width=8.5cm]{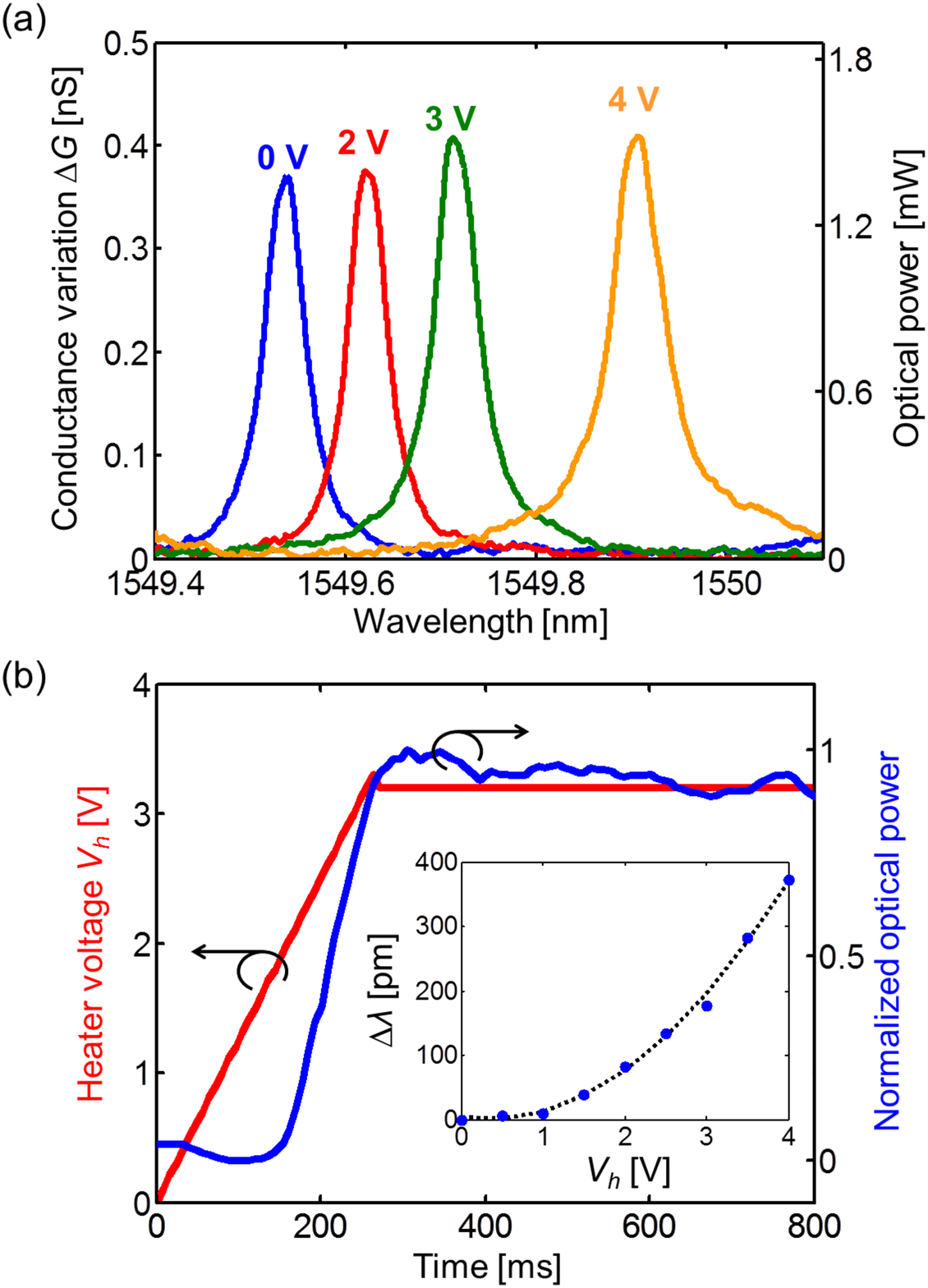}
\caption{Tuning of the microring resonant wavelength assisted by the CLIPP. (a) Light induced conductance variation $\Delta{G}$ measured by the CLIPP, and corresponding estimated optical power, as a function of wavelength, when the thermal actuator is off (blue line) and then switched on at $ V_{h} $ = 2, 3, 4 V (red, green, and orange lines). (b) Automated tuning of the resonator wavelength to that of an external laser, here detuned by about 230 pm, assisted by the CLIPP; the inset shows the wavelength shift $ \Delta\lambda $ measured by the CLIPP as the heater voltage $ V_{h} $ is increased from 0 to 4 V.}
\label{fig:Fig2}
\end{figure}

Read-out operations of the CLIPP electric signal, as well as microring control functionalities (such as wavelength tuning, locking, labeling and swapping), are performed by means of a custom microelectronic circuit \cite{Carminati_2014_Impedimetric}, whose schematic is shown in Fig. \ref{fig:Fig1}(a). One of the CLIPP electrodes is excited with a sinuosidal electrical signal, with frequency $ f_{e} $ and amplitude $ V_{e} $, whereas the current flow $ i_{e} $ at the other one is collected by means of a low noise amplifier. Then, the CLIPP signal is demodulated at frequency $ f_{e} $ to provide the electric conductance $ G $. The electric read-out frequency of the CLIPP is typically around $ f_{e} $ = 1 MHz in order to bypass the capacitances $C_A$ and access directly the waveguide conductance $G$, whereas the amplitude of the applied signal is usually $ V_{e} $ = 1 V, so that neither attenuation nor significant perturbation of the optical mode is induced, and without affecting the quality factor of the resonator \cite{Morichetti_2014_CLIPP}. Similarly, the integrated electronic circuit manages control operations of the microring, like wavelength tuning, locking, labeling and swapping, by using the dithering and feedback controller units reported in the schematic (see Sec. \ref{sec:tuning},  \ref{sec:locking}, and \ref{sec:swapping}).

The electronic read-out circuit is integrated in a CMOS chip (0.35 $ \mu $m process by AMS foundry) with 32 read-out channels, that is wire-bonded to the Si photonics chip hosting the microring resonator (fabricated by the James Watt Nanofabrication Centre at University of Glasgow) [Fig. \ref{fig:Fig1}(b)]. Both the electronic and the photonic chips are integrated onto the same printed circuit board. This photonic-electronic integrated platform enables to achieve not only improved CLIPP performances (lower noise, lower parasitics, better sensitivity, higher speed etc.) \cite{Carminati_2014_Impedimetric}, but also offers a simple and flexible system for the realization and control of large scale Si photonics integrated circuits, hosting several components to achieve complex systems-on-a-chip.  

\section{Tuning the microring resonant wavelength}
\label{sec:tuning}

Figure \ref{fig:Fig2}(a) shows the variations of conductance $\Delta{G}$ induced by the propagation of quasi-transverse electric (TE) polarized light in the resonator, measured by the CLIPP versus wavelength, when the thermal actuator is off (blue line), and then driven with voltage $ V_{h} $ = 2, 3, 4 V (red, green and orange lines). The microring has a linewidth of 51 pm (6.4 GHz), free-spectral-range of 1.115 nm (139.2 GHz), and quality factor of about 30000. Also, the corresponding optical power travelling in the resonator is provided on the rightmost vertical axis, as estimated from the conductance variations measured by the CLIPP \cite{Morichetti_2014_CLIPP}. Any spurious conductance change due to thermal cross-talk effects between CLIPP and heater is here negligible, being more than one order of magnitude smaller than $\Delta{G}$ induced by light at the low power levels utilized in this work.  

The effectiveness of the CLIPP to monitor the transfer function of the microring is exploited to automatically tune its resonant wavelength in order to overlap with that of an external laser [Fig. \ref{fig:Fig2}(b)]. The laser wavelength is initially red-shifted with respect to that of the resonator by about 230 pm (4.5 times the ring linewidth), then while the heater voltage $ V_{h} $ is automatically and continuously increased to shift the resonant wavelength (red line) the CLIPP simultaneously monitors the optical intensity stored in the cavity (blue line). The inset of Fig. \ref{fig:Fig2}(b) shows the CLIPP monitoring the microring resonant wavelength versus the voltage applied to the heater. The tuning process, here achieved in about 260 ms, terminates when the optical power measured by the CLIPP reaches its maximum value, here for $ V_{h} $ = 3.2 V, condition that occurs only when the resonator wavelength is aligned to that of the laser.    

\section{Locking the microring resonant wavelength}

Here we demonstrate feedback control of the microring by locking its resonant wavelength to that of an external laser. To this aim we utilize the CLIPP to monitor the optical intensity in the microring, and the thermal actuator to adjust its resonant wavelength based on a feedback error signal provided by the CLIPP. In particular, we employ a common dithering technique \cite{Padmaraju_2014_JLT}, according to which a small modulation signal is applied to the resonator, and then, by mixing it with the modulated intracavity optical intensity measured by the CLIPP, an error signal is extracted and used to drive the feedback loop [as shown in the schematic of Fig. \ref{fig:Fig1}(a)]. 

\label{sec:locking}
\subsection{Generation and read-out of the error signal}

Figure \ref{fig:Fig3}(a) shows the optical intensity in the resonator measured by the CLIPP, here low-pass filtered, when a sinusoidal dithering signal with frequency $ f_{d} $ = 160 Hz and amplitude $ V_{d} $ = 100 mV is applied to the heater of the resonator, as in the schematic of Fig. \ref{fig:Fig1}(a). In addition, a 2 V bias is applied to the heater in order to have the resonant wavelength overlapped to that of the laser.
 The CLIPP monitors also the corresponding error signal $ \varepsilon $ [Fig. \ref{fig:Fig3}(b), red line] by demodulating the resonant optical power at frequency $ f_{e} $+$ f_{d} $ = 1.00016 MHz. The error signal is zero at the resonant wavelength ($\lambda_{r}$ = 1549.54 nm) and maximum on the slope of the resonator ($\lambda_{r} \pm$ 20 pm). Also, thanks to the antysimmetric shape of $ \varepsilon $, no ambiguity is left to what direction $ \lambda_{r} $ is located. Here, the application of a dithering signal with 100 mV amplitude corresponds to a thermal fluctuation as low as $ \Delta{T} $ = 0.14 K (wavelength shift $ \Delta{\lambda} $ = 11 pm corresponding to about 20\% of the linewidth), that is in line with those used for tap detectors \cite{Padmaraju_2014_JLT} and does not affect the quality of the transmitted signal \cite{Padmaraju_2013_OpEx} . Though small, the amplitude of $ V_{d} $ can be further reduced to minimize the induced $\Delta{\lambda}$: in fact, with amplitudes of 50 mV and even 20 mV [Fig. \ref{fig:Fig3}(b), green and orange lines], corresponding respectively to $\Delta{T}$ = 0.07 K ($ \Delta{\lambda} $ = 5 pm, 10\% of the linewidth) and $\Delta{T}$ = 0.03 K ($ \Delta{\lambda} $ = 2 pm, 4\% of the linewidth), the error control signal is well above the noise level and can be used to drive the feedback loop. 

\subsection{Implementation of the control loop}

\begin{figure}[b]
\centering\includegraphics[width=9cm]{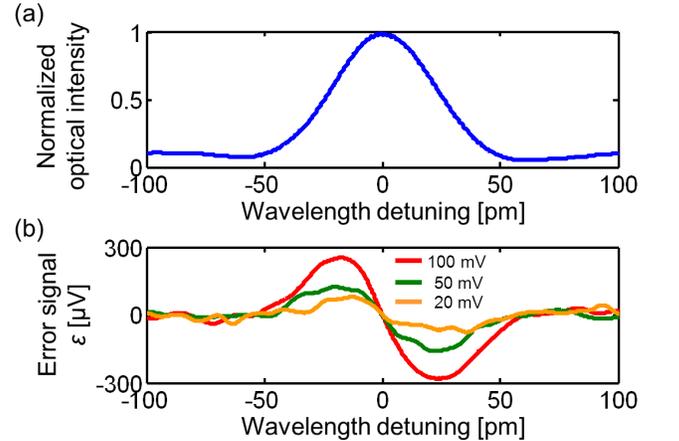}
\caption{Generation and read-out of the error signal of the feedback-controlled microring. (a) Normalized optical intensity measured by the CLIPP versus wavelength, with $ V_{e} $ = 1 V and $ f_{e} $ = 1 MHz, when a dithering signal with amplitude $ V_{d} $ = 100 mV and frequency $ f_{d} $ = 160 Hz is applied to the heater. (b) Error signal $ \varepsilon $ extracted by the CLIPP by further demodulating the optical power $ P $ at the frequency of the dithering signal $ f_{d} $ for dithering amplitudes $ V_{d} $ = 20, 50, 100 mV.}
\label{fig:Fig3}
\end{figure} 

In order to have the resonator wavelength $ \lambda_{r} $ continuously locked to that of an external laser $ \lambda_{l} $, the CLIPP monitors simultaneously and continuously the intracavity optical intensity and the level of the error signal $ \varepsilon $. As $ \lambda_{r} $ and $ \lambda_{l} $ drift apart from each other, the resonant optical power drops and the error signal deviates from zero. Consequently, the voltage applied to the heater is updated with an increment proportional to $ \varepsilon $, its sign indicating the direction to follow (heating or cooling), and thus restoring the alignment between $ \lambda_{r} $ and $ \lambda_{l} $. 

The feedback loop is implemented by means of an integral controller, whose gain $ k $ depends on the magnitude of $ \varepsilon $ with respect to the wavelength detuning $ \Delta\lambda $ [Fig. \ref{fig:Fig3}(b)], and on the wavelength shift that the heater can achieve [inset of Fig. \ref{fig:Fig2}(b)]. According to our model, $ k $ should be sufficiently high to achieve fast response of the feedback loop, but, at the same time, low enough to guarantee stability of the system. As an example, considering that $\Delta\varepsilon $/$\Delta\lambda $ = 15 $ \mu $V/pm around the resonant wavelength when $ V_{d} $ = 100 mV, and that the heater shifts $ \lambda_{r} $ by about 95 pm/V around 2 V bias, a controller gain around $ k $ = 10000 is sufficient to provide a loop response as low as 50 ms, while maintaining the system stability according to the Bode criterion. Here, the controller is implemented by means of a programmable digital platform (FPGA), thus allowing more speed and flexibility in setting the controller parameters with respect to computer-assisted architectures.

\subsection{Testing the control loop}

\begin{figure}[t]
\centering\includegraphics[width=8.4cm]{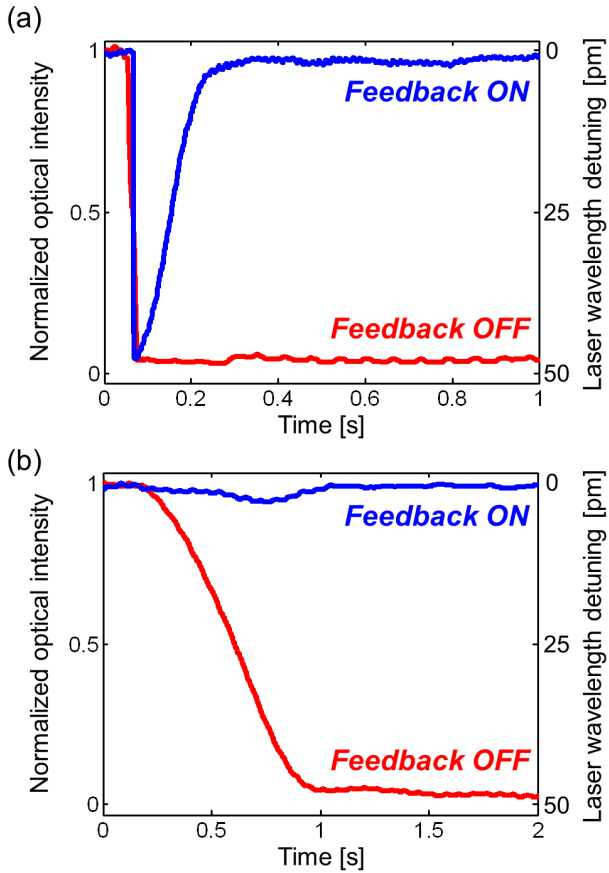}
\caption{Locking the microring resonant wavelength to that of an external laser $ \lambda_{l} $ assisted by the CLIPP. Normalized optical intensity in the resonator measured by the CLIPP when the feedback control is on (blue lines) and off (red lines) in presence of (a) an istantaneous wavelength shift and (b) a continuous wavelength sweep (here occurring in about 700 ms) of the external laser by 50 pm (corresponding to 98\% of the resonator linewidth, that is the same of a temperature variation of 0.7 K).}
\label{fig:Fig4}
\end{figure}

\begin{figure*}[t]
\centering\includegraphics[width=14cm]{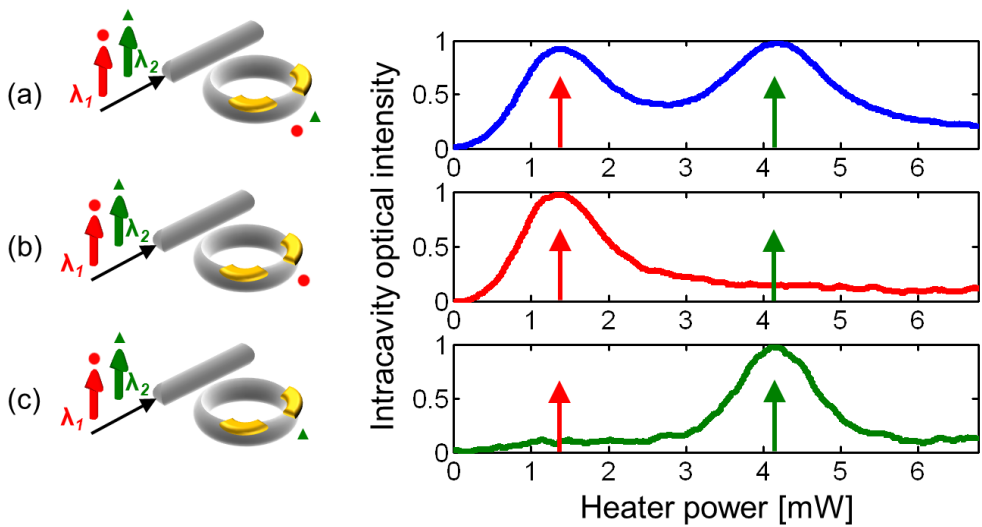}
\caption{Swapping the resonator wavelength between two optical signals at wavelengths $ \lambda_{1} $ (red arrow) and $ \lambda_{2} $ (green arrow) injected in the microring. A weak modulation tone with depth 2$ \% $ is added to label each of the optical carriers: the tone centered on $ \lambda_{1} $ has frequency $ f_{1} $ = 10 kHz (labeled with a red circle), whereas the tone around   $ \lambda_{2} $ has frequency $ f_{2} $ = 11 kHz (labeled with a green triangle). The optical intensity measured by the CLIPP in the resonator is reported as a function of the electrical power dissipated on the heater when the CLIPP signal is demodulated at frequencies (a) $ f_{e} $ = 1 MHz, (b) $ f_{e} $+$ f_{1} $ = 1.010 MHz, (c) $ f_{e} $+$ f_{2} $ = 1.011 MHz. When demodulating the CLIPP signal at frequency $ f_{e} $, the signals $ \lambda_{1} $ and $ \lambda_{2}$ are indistinguishable; in constrast, when read-out operations are performed at frequency $ f_{e} $+$ f_{1} $ ($ f_{e} $+$ f_{2} $) the CLIPP is able to identify distinctively $ \lambda_{1} $ ($ \lambda_{2} $).}
\label{fig:Fig5}
\end{figure*}

The feedback loop is here tested against external fluctuations of the laser wavelength. Figure \ref{fig:Fig4} shows the optical intensity in the microring, measured by the CLIPP as a function of time, when the feeback routine is on (blue lines) and off (red lines), in presence of a detuning of $ \lambda_{l} $ by 50 pm (about the same of the resonator linewidth, and corresponding to a temperature variation of 0.7 K), i.e. approximately at the edges of its full linewidth. In Fig. \ref{fig:Fig4}(a) an instantaneous wavelength shift is imposed by the laser: the optical intensity in the resonator drops as $ \lambda_{l} $ is changed, but then is rapidly brought back (here in about 150 ms) to its initial level (where $ \lambda_{l} $ and $ \lambda_{r} $ overlap) by the feedback loop (blue line). 
In Fig. \ref{fig:Fig4}(b), instead, the feedback loop compensates for a continuous wavelength sweep of the laser: no significant variation in the resonant optical intensity is observed (blue line) along the entire wavelength detuning (here occurring in about 700 ms), having $ \lambda_{l} $ and $ \lambda_{r} $ always locked one to another by the feedback control. The beneficial effect of the feedback loop is confirmed when the control is switched off (red lines): in fact in both cases, as $ \lambda_{l} $ shifts, the optical intensity in the resonator drops and is never compensated, having the laser wavelength completely outside of the resonator linewidth. 

Although the speed of the control loop here achieved is enough to well counteract most of the drifts (thermal and others) experienced by Si photonics microrings, it can be significantly increased either with suitable control laws (such as proportional or proportional-integral) or by enlarging the CLIPP read-out bandwidth. For instance, while keeping the current CLIPP bandwidth (around 1 kHz), if a proportional controller with gain of few thousands is utilized, a loop with time constant down to hundreds of $ \mu $s is achieved, yet at the price of larger voltage variations at the thermal actuator. On the other side, while maintaining an integral controller, the loop frequency can be increased by using a larger CLIPP bandwidth, yet at the price of larger noise (though typically depending on the square root of the bandwidth).

\section{Swapping the microring resonant wavelength}
\label{sec:swapping}

Although the CLIPP is in general a broadband light observer, here we show that it is able to monitor and discriminate simultaneously optical signals at different wavelengths. To this aim we inject in the microring two different signals with wavelengths $ \lambda_{1} $ = 1549.59 nm and $ \lambda_{2} $ = $ \lambda_{1} $ + 120 pm, labeled respectively with red and green arrows in Fig. \ref{fig:Fig5}. A weak modulation tone, with depth 2$ \% $, is added to label each of the optical carriers by means of an external modulator. The tone centered around $ \lambda_{1} $ has frequency $ f_{1} $ = 10 kHz, whereas the other one, that is centered around $ \lambda_{2} $, has frequency $ f_{2} $ = 11 kHz (the tones are labeled respectively with red circle and green triangle in Fig. \ref{fig:Fig5}). The optical intensity observed in the resonator by the CLIPP is shown in Fig. \ref{fig:Fig5} as a function of the electrical power dissipated by the heater. When the CLIPP electrical signal is demodulated at the usual frequency $ f_{e} $ = 1 MHz, the intracavity optical intensity reports two peaks associated to the transmission of both signals $ \lambda_{1} $ and $ \lambda_{2} $ [blue line in Fig. \ref{fig:Fig5}(a)], that are therefore indistinguishable. Vice versa, if the read-out operations are performed at frequency $ f_{e} $+$ f_{1} $ = 1.010 MHz, the CLIPP is able to identify the signal at $ \lambda_{1} $, where the modulation tone at frequency $ f_{1} $ is added. In fact, as shown in Fig. \ref{fig:Fig5}(b) with red line, the resonant peak measured by the CLIPP is centered on the first signal (red arrow), whereas no evidence of $ \lambda_{2} $ is found (green arrow). Similarly, if the electric signal is demodulated at frequency $ f_{e} $+$ f_{2} $ = 1.011 MHz, the measured resonant peak is centered on the second signal (green arrow), that is identified by the CLIPP as wavelength $ \lambda_{2} $. It is straightforward now to tune and lock the microring resonant wavelength to that of the first signal $ \lambda_{1} $ ($ \lambda_{2} $), and then easily swap it to that of the second signal $ \lambda_{2} $ ($ \lambda_{1} $). Furthermore, it is worth noticing that wavelength swapping operations can be performed by the CLIPP simultaneously on an arbitrarily large number of wavelengths by using several modulation tones and by suitably demodulating the different signals, on signals at the same wavelength but different light polarizations \cite{Morichetti_2014_CLIPP}, and on higher order modes in waveguides exploiting mode-division multiplexing schemes \cite{Luo_NatComm_2013}. 

\section{Conclusion}

We demonstrated non-invasive monitoring and advanced control functionalities in Si photonics microring resonators, assisted by a fully transparent light detector directly integrated inside the cavity. Through a CMOS microelectronic circuit \cite{Carminati_2014_Impedimetric} bridged to the Si photonics chip, the resonant wavelength of the microring is automatically tuned and locked against wavelength drifts of the optical signal. The non-invasive nature of the CLIPP enables real-time inspection of the intra-cavity light intensity without affecting the quality factor of the resonator. Furthermore, the CLIPP is able to monitor and discriminate signals, at different wavelengths, simultaneously resonating in the microring, thus enabling wavelength swapping operations. In the fabricated devices, feedback control is achieved through thermal actuators, but the proposed tuning and locking schemes, and the CLIPP concept itself, can be adopted with any available actuator technology.

Finally, the compactness and the scalability of CMOS electronics to multi-channel read-out systems and the transparency of the CLIPP enable not only multipoint on-chip monitoring \cite{Morichetti_2014_CLIPP}, but also the realization and control of photonic integrated circuits hosting many components and complex systems-on-a-chip.


\section*{Acknowledgments} 
The authors are grateful to the staff of the James Watt Nanofabrication Centre (JWNC) at University of Glasgow for the fabrication of the Si photonics sample, and to Giovanni Bellotti from Politecnico di Milano for support during the measurements.

\section*{Funding} 
This work was supported by the european project BBOI (Breaking the Barriers of Optical Integration, website: www.bboi.eu) of the 7$^{th}$ EU Framework Program.

\end{document}